\documentclass[12pt]{article}

\usepackage{amsmath,amssymb, mathtools}
\usepackage{graphicx}
\usepackage{caption}
\usepackage{color}
\usepackage{dcolumn}
\usepackage{bm}
\usepackage{float}
\usepackage{hyperref} 
\usepackage{apacite}
\hypersetup{colorlinks = true, citecolor = black, urlcolor = blue}
\usepackage{silence} 
\definecolor{background-color}{gray}{0.98}

\DeclareMathOperator{\argmax}{argmax}

\newcommand{\RD}{\mbox{RD}}
\newcommand{\PAC}{\mbox{PAC}}
\newcommand{\cond}{\,|\,}

\newcommand{\bx}{\boldsymbol{x}}
\newcommand{\bS}{\boldsymbol{S}}
\newcommand{\bX}{\boldsymbol{X}}

\newcommand{\bmu}{\boldsymbol{\mu}}
\newcommand{\bSigma}{\boldsymbol{\Sigma}}

\newcommand{\hbmu}{\hat{\boldsymbol{\mu}}}
\newcommand{\hbSigma}{\hat{\boldsymbol{\Sigma}}}
\newcommand{\hpi}{\hat{\pi}}
\newcommand{\hf}{\hat{f}}
\newcommand{\hg}{\hat{g}}

\providecommand{\red}[1]{\textcolor{black}{#1}}

\title{\bf Robust discriminant analysis}

\author{Mia Hubert\thanks{Department of 
    Mathematics, KU Leuven, Belgium}, 
    Jakob Raymaekers\thanks{Department of 
    Mathematics, University of Antwerp}, and 
    Peter J. Rousseeuw\footnotemark[1]\\
		\vspace{3mm}}

\date{August 28, 2024}
\begin{document}
\maketitle

\begin{center}
\vspace{5mm}

\large
Accepted for publication in WIREs Computational Statistics\\
(Wiley Interdisciplinary Reviews)\\

\hfill \break
\thanks

\vspace{5mm}

{Abstract}
\end{center}

\large
Discriminant analysis (DA) is one of the most popular 
methods for classification due to its conceptual 
simplicity, low computational cost, and often solid 
performance. In its standard form, DA uses the arithmetic 
mean and sample covariance matrix to estimate the center 
and scatter of each class. We discuss and illustrate how 
this makes standard DA very sensitive to suspicious data 
points, such as outliers and mislabeled cases. We then 
present an overview of techniques for robust DA, which 
are more reliable in the presence of deviating cases. In 
particular, we review DA based on robust estimates of 
location and scatter, along with graphical diagnostic 
tools for visualizing the results of DA.

\clearpage

\renewcommand{\baselinestretch}{1.5}
\normalsize

\newpage
\section*{\sffamily \Large INTRODUCTION} 

Discriminant analysis (DA) is a supervised classification 
technique for multivariate data. It involves 
constructing rules that describe the 
separation between known classes (groups). These rules then 
allow to classify new cases into one of the classes.

Standard discriminant analysis relies on the empirical mean 
and covariance matrix of each class, making it very sensitive 
to outlying cases. The training data can be contaminated by 
\textit{measurement noise}, which occurs when the cases have 
deviating measurements. \textit{Label noise}, also called 
mislabeling, occurs when instances in the training data have 
been given a wrong label, so their recorded label differs 
from their true one. 

Robust discriminant \red{analysis} relies on robust estimates of the center 
and shape of each class. These can be obtained by the Minimum 
Covariance Determinant (MCD) estimator \cite{Rousseeuw:MCD}. 
We also illustrate several graphical tools to display various 
aspects of the classification results, among which the 
silhouette plot \red{\cite{Silh1987}} and the class map \red{\cite{Raymaekers:ClassMap}}. 

\section*{\sffamily \Large CLASSICAL DISCRIMINANT ANALYSIS}

\subsection*{\sffamily \large Methodology}

In the classification setting one assumes that training 
data are sampled from a $(p+1)$-dimensional random 
variable $(\bX,\cal{G})$ with discrete response variable 
$\cal{G}$ taking values $g = 1,\ldots, G$. The 
\textit{prior probability} of class $g$ is defined as 
$\pi_g = P({\cal G} = g)$.  
It is assumed that $\bX$ has a density $f$ which satisfies  
the mixture model: 
$$f(\bx) = \sum_{g = 1}^{G}{\red{\pi_g} f_g(\bx)}$$ with $f_g$ the 
conditional density of $\bX$ given class $g$.

The Bayes classification rule \red{maximizes the probability
to guess correctly the value of the variable $\cal{G}$.}
It assigns a new case $\bx$ to 
the class $g$ with the highest posterior probability 
$P(g \cond \bx)$. Since $P( g \cond \bx)$ is proportional 
to $P( \bx \cond g) \pi_g$, this is equivalent to 
allocating $\bx$ to the class with highest $\pi_g f_g(\bx)$.

Assuming $\bX \cond g \sim N_p(\bmu_g, \bSigma_g)$ 
one obtains {\it quadratic discriminant analysis} 
(QDA), which assigns $\bx$ to the class~$g$ 
with highest \textit{quadratic discriminant score} 
\begin{equation} \label{eq:QDA}
ds(\bx, \bmu_g, \bSigma_g, \pi_g) = 
  -\frac{1}{2} \ln|\bSigma_g| - \frac{1}{2} 
  (\bx- \bmu_g)' \bSigma_g^{-1} (\bx- \bmu_g)
  + \ln(\pi_g) \,.
\end{equation}
When all covariance matrices $\bSigma_g$ are equal to a 
common covariance $\bSigma$, the discriminant scores can
be simplified to
\begin{equation} \label{eq:LDA}
ds^L(\bx, \bmu_g, \bSigma, \pi_g) =
\bmu_g' \bSigma^{-1} \bx  - 
\frac{1}{2} \bmu_g' \bSigma^{-1} \bmu_g + \ln(\pi_g) 
\end{equation}
because $|\bSigma_g|$ and $\bx' \bSigma_g^{-1} \bx$ 
no longer depend on $g$\,. The scores \eqref{eq:LDA}
are linear in $\bx$, so the resulting method is
called {\it linear discriminant analysis} (LDA).

In practice one has sampled training data 
$(\bx_i, g_i)$ for $i=1,\ldots,n$\,, with 
\textit{class labels} $g_i \in \{1, \ldots, G\}$\,. 
The number of cases in class $g$ is denoted as $n_g$. 
\red{It is assumed that the data from class $g_i$ are sampled from $N_p(\bmu_g, \bSigma_g)$.}

Classical Quadratic Discriminant Analysis (CQDA) 
estimates the center and shape of each class $g$ by 
its arithmetic mean $\hat{\bmu}_g = \bar{\bx}_g$ and 
empirical covariance matrix $\hat{\bSigma}_g=\bS_g$. 
The estimated prior probabilities 
$\hat{\pi}_{g,C} = n_g/n$ of classical QDA are 
proportional to the class sizes. A case $\bx$ is 
then assigned to the class $g$ with highest 
$ds(\bx,\bar{\bx}_g,\bS_g,\hat{\pi}_{g,C})$.
Note that each class should contain at least $p+1$ 
cases, because otherwise $\bS_g$ becomes singular and 
the quadratic discriminant scores cannot be computed. 
In Classical Linear Discriminant Analysis (CLDA) the 
presumed common covariance matrix $\bSigma$ is instead
estimated by the pooled empirical covariance matrix 
$\bS_p = \frac{1}{n - G} \sum_{g=1}^G (n_g-1) \bS_g$.
More information on classical DA can be found in  
\cite{mclachlan2004discriminant}.

\subsection*{\sffamily \large Sensitivity to outliers}

\begin{figure}[!ht]
	\centering
	\includegraphics{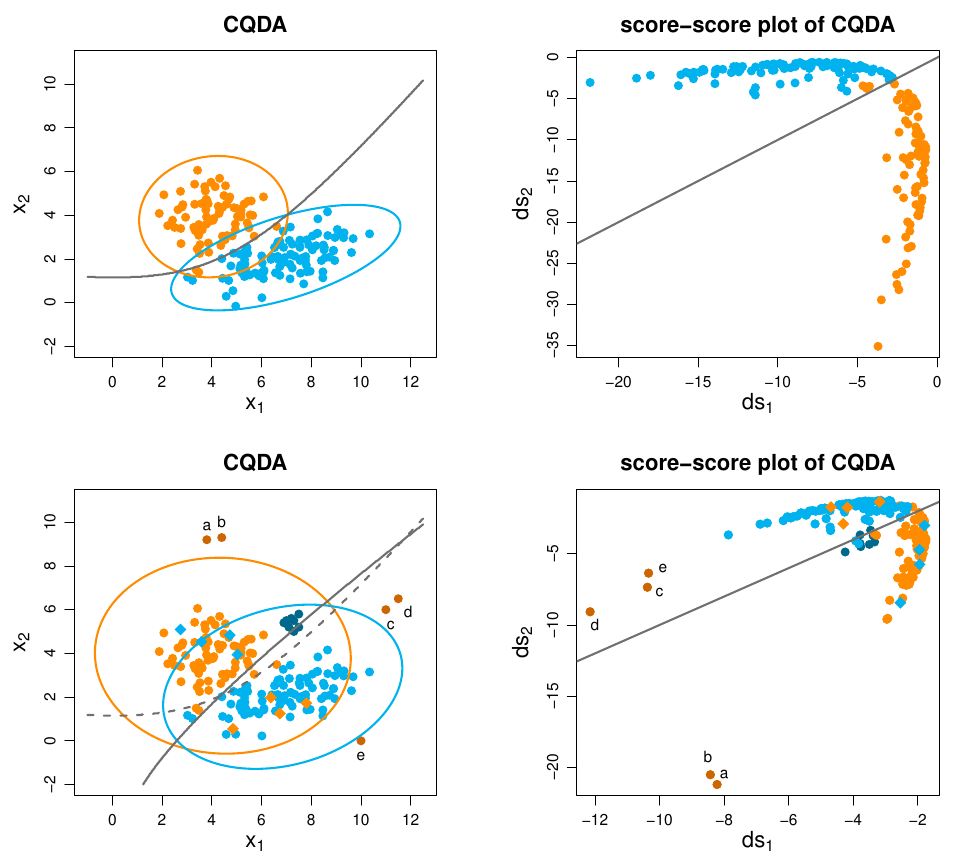}
 	\caption{The effect of label and measurement noise on CQDA. Top row: two uncontaminated classes with the CQDA decision boundary in the left figure, and the corresponding score-score plot on the right with the identity line separating the groups. Bottom row: the same plots for contaminated data. Mislabeled cases are depicted as  diamonds, and outliers have a darker color. In the bottom left figure the dashed curve is the classification boundary from the top left figure.}
\label{fig:noisyCQDA}
\end{figure}

Although CQDA and CLDA are popular classification methods, 
it has been known for quite some time that they are very 
sensitive to outliers and mislabeled cases 
\cite{Campbell1978}. To illustrate this we generated 
data from two bivariate normal distributions, yielding
the two classes in the top left panel of 
Figure~\ref{fig:noisyCQDA}. There are 80 cases 
of class~1 (orange) and 100 cases of class~2 (blue).
We then applied CQDA. \red{The blue and orange curves are the so called tolerance ellipses. 
They} correspond to (hypothetical) points $\bx$ whose 
Mahalanobis distance 
\begin{equation}
\label{eq:mahalan}
  \text{MD}(\bx,\bar{\bx}_g,\bS_g) = 
  \sqrt{(\bx- \bar{\bx}_g)' \bS_g^{-1} 
	(\bx- \bar{\bx}_g)}
\end{equation}
equals {\scriptsize $\sqrt{\chi^2_{2,0.99}}$}\,, the 
square root of the 0.99 quantile of the $\chi^2$ 
distribution with $p=2$ degrees of freedom. They 
visualize the empirical covariance matrices and fit 
the data nicely. \red{When data are normally distributed in each class, we would expect only 1\% of the cases to lie outside their respective tolerance ellipse. This is a quite conventional choice which avoids flagging too many regular observations as outliers.}  The gray curve is the quadratic 
decision boundary obtained by CQDA. It separates 
the classes quite well, misclassifying 
only three orange points into the blue class. It 
is natural to have some misclassification, as the 
classes overlap a bit. 

The classification can also be visualized by means 
of a score-score plot, inspired by the depth-depth 
plot of \red{\cite{Liu:Depth}} and the distance-distance 
plot introduced in \cite{Hubert:MFC} for the 
classification of functional data. 
It displays the discriminant scores of class~2 
versus those of class~1, i.e.
$ds_2(\bx_i) = ds(\bx_i,\hat{\bmu}_2,
\hat{\bSigma}_2,\hat{\pi}_{2})$ versus 
$ds_1(\bx_i) = ds(\bx_i,\hat{\bmu}_1,
\hat{\bSigma}_1,\hat{\pi}_{1})$, for all points 
$\bx_i$ in the training data. 
This yields the plot in the top right panel of
Figure~\ref{fig:noisyCQDA}, based on the CQDA fit. 
The gray identity line $ds_1=ds_2$ corresponds 
with the CQDA separation boundary. The few 
misclassified cases from class~1 are easily spotted.

In the bottom left panel we have introduced label 
noise, by replacing the label of four cases of each 
class by the label of the other class. They are 
depicted as diamonds. We also replaced five cases 
from class~1 by outlying points in dark orange, and 
eight cases of class~2 by outliers in dark blue, 
thereby introducing measurement noise. 
The substantially larger ellipses and the modified 
decision boundary show how much CQDA is affected. 
For comparison, the dashed curve in the left figure 
is the decision boundary that was computed from
the uncontaminated data in the panel above it.
If the model was trained on the outlier-free data, 
the method would assign test data in the position
of the dark blue cluster to the orange class.
When this cluster is included in the training set, 
part of it is assigned to the blue class.

\section*{\sffamily \Large ROBUST DISCRIMINANT ANALYSIS}

\subsection*{\sffamily \large Methodology}
Robust discriminant rules can be obtained by plugging 
in robust estimators of location and scatter of each 
class into~\eqref{eq:QDA} or~\eqref{eq:LDA}.
For this purpose \cite{Chork:DA-MVE} used the Minimum 
Volume Ellipsoid estimator \cite{Rousseeuw:LMS}, 
whereas \cite{He:Discrim} and \cite{Croux:Discrim} 
employed multivariate S-estimators in LDA.

In \cite{Hubert:Discrim} the QDA and LDA methods 
were robustified by inserting the Minimum Covariance 
Determinant (MCD) estimates of each class. The MCD
looks for a subset of the class which contains a 
fraction $\alpha$ of its cases, such that the sample 
covariance matrix of that subset has the smallest 
possible determinant. Its location is then the mean
of that subset, and its scatter matrix is a multiple 
of that sample covariance matrix. The MCD was
proposed in \cite{Rousseeuw:LMS} and 
\cite{Rousseeuw:MCD}. \red{The estimator is affine equivariant 
which implies that the data may be rotated, translated or rescaled (e.g., through a change of measurement
units) without affecting the outlier detection diagnostics. Its influence function is bounded. This indicates  that a
small fraction of outliers has a limited effect on the estimate.} 
The robustness of the MCD \red{towards a large proportion of outliers} depends on 
the choice of $0.5 \leqslant \alpha < 1$ which 
reflects a lower bound on the proportion of cases in 
each class that are presumed to be outlier-free. 
Typical choices are $\alpha = 0.5$ when many outliers 
might be present in the data, and $\alpha = 0.75$ for
a less extreme fraction of contamination. 
In order to increase the statistical efficiency of
the MCD, it is standard practice to follow it by a 
reweighting step.  For a \red{more detailed} review of its
properties see \cite{Hubert:WIRE-MCD2}.

The MCD estimator can be computed by the 
FastMCD algorithm \cite{Rousseeuw:FastMCD} which 
performs well at small to moderate data sizes. 
For larger data sets the deterministic algorithm 
DetMCD \cite{Hubert:DetMCD} turns out to be faster 
than FastMCD and less sensitive to concentrated 
contamination. Both algorithms are available in
\textsf{R} \red{\cite{RCT2023} and yield the reweighted estimates}. The \texttt{robustbase} package 
\cite{robustbase} contains the function 
\texttt{covMcd}, and the \texttt{rrcov} package
\cite{rrcov} has the function \texttt{CovMcd}.
For calculations in real time the RT-DetMCD 
algorithm was developed \cite{DeKetelaere:RTDetMCD}, 
which uses parallel computation.

The full description of the robust DA approach
goes as follows. For each class $g$ robust 
estimates $\hbmu_{g,\text{R}}$ and 
$\hbSigma_{g,\text{R}}$ are computed, where the 
subscript R stands for robust. These estimates 
can for instance be obtained by the MCD method.
Next, the {\it robust distance} of every case 
$\bx_i$ to its own class $g_i$ is given by
\begin{equation} \label{eq:RD}
	\RD_{i,g_i} = \RD(\bx_i,\hbmu_{g_i,\text{R}}, 
	\hbSigma_{g_i,\text{R}}) = 
	\sqrt{{(\bx_i- \hbmu_{g_i,\text{R}})' 
	\hbSigma_{g_i,\text{R}}^{-1} 
	(\bx_i- \hbmu_{g_i,\text{R}})}}\;.
\end{equation}  
A case $(\bx_i,g_i)$ can then be flagged as an 
outlier when its robust distance from its own 
class is too large, i.e.\ when $\RD_{i,g_i} >$ 
{\scriptsize $\sqrt{\chi^2_{p, 0.99}}$}. This 
may happen as a result of measurement noise 
(outlying $\bx_i$) or label noise (incorrect $g_i$). 
These flagged cases are then removed from the 
calculation of the prior probabilities, yielding 
robust estimates  
$\red{\hat{\pi}_{g,\text{R}}} = \tilde{n}_g/\tilde{n}$ 
where $\tilde{n}_g$ denotes the number of 
unflagged cases in class $g$ and 
$\tilde{n} = \sum_{g=1}^{G}{\tilde{n}_g}$\,. 
Inserting this \red{$\hat{\pi}_{g,\text{R}}$} together 
with $\hbmu_{g,\text{R}}$ and 
$\hbSigma_{g,\text{R}}$ into~\eqref{eq:QDA} 
then yields robust discriminant 
scores, which are used to assign cases as before.
This approach was taken in \cite{Hubert:Discrim}
using the FastMCD algorithm, but one could also
apply the DetMCD algorithm. In 
\cite{Vranckx:RT-RQDA} the RT-DetMCD estimates
were used.

\subsection*{\sffamily \large Robustness to outliers}

\begin{figure}[!ht]
\centering
\includegraphics{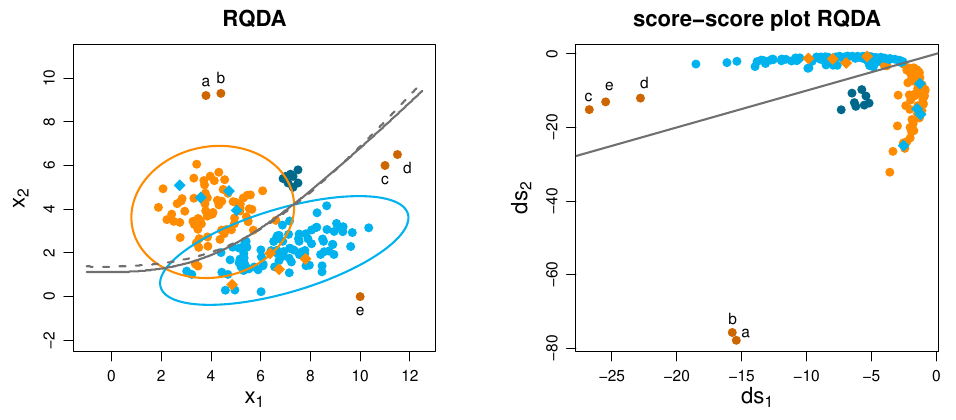}
\caption{The effect of label and measurement noise on 
robust QDA. The solid curve in the left panel is the 
robust decision boundary computed on the contaminated 
data, whereas the dashed curve is that of RQDA on the 
original uncontaminated data in 
Figure~\ref{fig:noisyCQDA}.}
\label{fig:noisyRQDA}
\end{figure}

Applying RQDA with DetMCD \red{and $\alpha = 0.75$} to the contaminated data
in the lower left panel of Figure~\ref{fig:noisyCQDA} 
yields the left panel of Figure~\ref{fig:noisyRQDA}. 
The new decision boundary is far less affected by 
contamination than the CQDA boundary was. The first 
two columns of Table~\ref{tab:RQDA_conting} contain 
the resulting confusion matrix, in which 10 cases 
with label 1 and 12 cases with label 2 are 
misclassified, hence the accuracy 
is $(70+88)/\red{180} = 87.8\%$.

\begin{table}[!ht]
\centering
\begin{tabular}{ccccccc}
 & \multicolumn{5}{c}{predicted}  \\
 \cline{2-7}
  & class 1 & class 2 &  & class 1 & class 2 & outliers\\
 \cline{2-3}  \cline{5-7} 
 class 1  & 70& 10 & & 68 & 7 & 5 \\ 
 class 2 & 12 & 88 & & 6 & 88 & 6 \\
 \hline
\end{tabular}
\caption{RQDA confusion matrices of the simulated data in Figure~\ref{fig:noisyRQDA}. The matrix formed by the first 
two columns considers only two predicted classes, whereas 
the last three columns form a table that includes an 
outlier class.}
\label{tab:RQDA_conting}
\end{table}

This accuracy is based on all the cases, as the standard 
DA rule assigns any training case and any test case to 
one of the known classes. This might however be 
inappropriate for a case that deviates strongly from all 
the given classes, for example when it belongs to a class 
that was not present in the training data or when it 
contains much measurement noise. A case whose robust 
distance~\eqref{eq:RD} to all $G$ classes exceeds the
cutoff can be considered an {\it overall outlier}, and 
assigned to an outlier class. 

In our example the dark blue cluster is now mostly 
assigned to the outlier class, as most of its cases 
fall outside the tolerance ellipses of both classes. 
The same holds for all the outlying cases of class 1. 
The resulting confusion matrix is presented in the last 
three columns of Table~\ref{tab:RQDA_conting}. 

\begin{figure}[!ht]
\centering
\includegraphics[width=8cm]{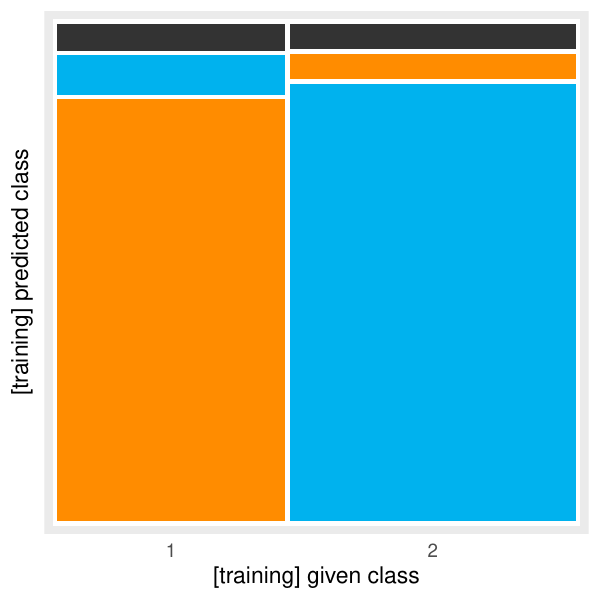}
\caption{Stacked plot of the data in 
Figure~\ref{fig:noisyRQDA}, with the outlier class 
in dark gray.}
\label{fig:RQDA_mosaic}
\end{figure}

This extended confusion matrix is represented
graphically in the stacked mosaic plot of 
Figure~\ref{fig:RQDA_mosaic}. The classes
are represented by their colors, with the outlier 
class in dark gray. The given classes are on the 
horizontal axis, and the predicted labels are on 
the vertical axis. The area of each rectangle is 
proportional to the number of objects in it.

The accuracy can now be evaluated without the 
outliers, leading to the higher value 
$(68+88)/(180-11) = 92.3\%$. 
Although this accuracy is a better summary of the 
performance of the classifier, it still includes 
the mislabeled cases from both classes. These 
cases are well classified according to their true 
label, but this is masked as their given label is 
wrong. In the next section we describe how one 
can distinguish between mislabeled cases and 
misclassified cases.    

\section*{\sffamily 
   \Large VISUALIZING CLASSIFICATION RESULTS}

In the previous section we saw scatterplots of 
data with a classification boundary added to it, but
these are only convenient as a graphical tool for 
bivariate data. Moreover, the score-score plot is 
limited to two classes ($G=2$). In this section the 
general setting is considered, with $p$-variate data 
($p \geqslant 2$) and $G \geqslant 2$ classes. 
The goal is to visualize the result of a discriminant 
analysis to gain insight into various features of the 
classification, including the quality of the fit, 
potential differences between the classes, and the 
presence of hard-to-classify points and outliers.

Discriminant analysis assigns each case to the class 
with the highest estimated posterior probability.
For a case $\bx_i$ the estimated posterior 
probability of a class $g$ is given by 
$$\widehat{P}( g \cond \bx_i) =  
  \frac{\hpi_g \hf_g(\bx)}
  {\sum_{g=1}^G{\hpi_g \hf_g(\bx_i)}},$$
where $\hf_g(\bx_i) = \phi(\bx_i,\hbmu_{g,\text{R}}, 
\hbSigma_{g,\text{R}})$ is the density of the 
multivariate normal distribution with estimated 
location and covariance matrix. By the maximum a 
posteriori classification rule, the class predicted 
for $\bx_i$ is then 
$\hg_i \coloneqq \argmax_{g\in \{1, \ldots, G\}}
{\widehat{P}( g \cond \bx_i)}$\,.
 
Now suppose case $\bx_i$ has a given class $g_i$ 
in the training data. The question is how well 
the classification agrees with this given label. 
In order to quantify 
this, one first defines the highest 
$\widehat{P}( g \cond \bx_i)$ attained by a 
class {\it different from} $g_i$ as
\begin{equation}\label{eq:altclass}
   \tilde{P}(\bx_i) = \max\{\widehat{P}
   ( g \cond \bx_i)\,;\,g \neq g_i\}\;.
\end{equation}
Clearly, case $\bx_i$ is misclassified when 
$\tilde{P}(\bx_i) > \widehat{P}( g_i \cond \bx_i)$, 
whereas 
$\tilde{P}(\bx_i) < \widehat{P}( g_i \cond \bx_i)$ 
implies that the classifier agrees with the given 
label $g_i$. From $\tilde{P}(\bx_i)$ one can compute 
the conditional posterior {\bf P}robability of the 
best {\bf A}lternative {\bf C}lass (PAC) as
\begin{equation}\label{eq:PAC}
   \PAC(\bx_i) = \frac{ \tilde{P}(\bx_i)}
   {\widehat{P}( g_i \cond \bx_i) + 
   \tilde{P}(\bx_i)}\;\;.
\end{equation} 
Note that $\PAC(\bx_i) > 0.5$ indicates that the 
classifier prefers to put $\bx_i$ in a class different 
from $g_i$\, whereas $\PAC(\bx_i) < 0.5$ means that 
the classifier agrees with the given label. 
The value of $\PAC(\bx_i)$ can \red{be} interpreted as a 
continuous measure of how well the given label $g_i$ 
agrees with the classifier. It ranges from 0, 
indicating a perfect fit, to 1, which says that
a class different from $g_i$ fits perfectly. 
In this sense it is similar to an absolute residual 
in regression analysis, where smaller is better too.

The silhouettes display for cluster analysis
\cite{Silh1987} displays for each case how well
it agrees with the cluster it was put in.
This is a similar notion to the PAC, but the scale 
is different, because for silhouettes the best fit 
corresponds to the value 1 and the worst fit to -1.
In order to stay with the existing convention, the
silhouette width of case $\bx_i$ in discriminant
analysis is defined as
\begin{equation}\label{eq:SI}
    s(\bx_i) = 1 - 2 \, \PAC(\bx_i)
\end{equation}
\cite{Raymaekers:Silhouettes}.
This ranges from -1 to 1, and high values indicate 
a good agreement between given label and predicted 
label, whereas low values indicate disagreement. 

\begin{figure}[!ht]
\centering
\includegraphics[width=0.8\linewidth]
  {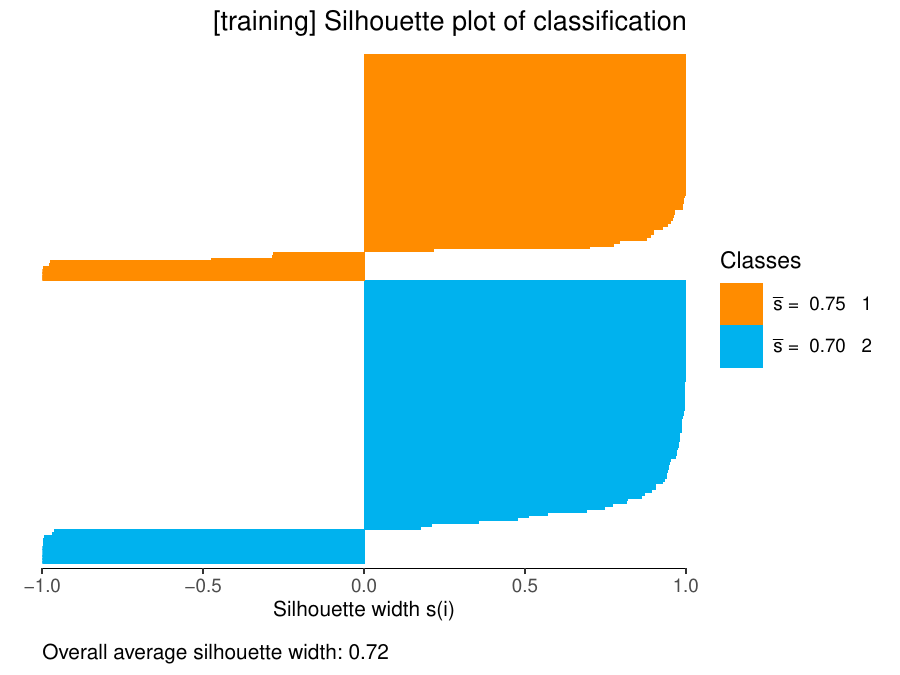} 
\caption{RQDA silhouette plot of the bivariate 
         data from Figure~\ref{fig:noisyRQDA}.}
\label{fig:RQDA_silh}
\end{figure}

Figure \ref{fig:RQDA_silh} displays the silhouette 
plot of RQDA on the bivariate data from 
Figure~\ref{fig:noisyRQDA}. 
The silhouette width is plotted in the horizontal
direction here, and it is often close to 1, so most 
labels agree with the classification by RQDA.
The overall average silhouette width of 0.72 is 
quite high. Class~1 attains an average silhouette 
width of 0.75, which indicates that it is a bit
better classified than class~2 which attains  
0.70\,. Both classes contain a few cases with
silhouette width close to -1. The given label of 
these cases does not match the predicted label.
This is understandable because we deliberately
mislabeled some cases in these data, as is clearly
visible in the bottom left panel of
Figure~\ref{fig:noisyCQDA} and in the left panel of
Figure~\ref{fig:noisyRQDA}. 

Whereas the silhouette plot for RQDA conveys more
insight about the data than a confusion matrix, one 
can extract even more information by plotting the 
PAC against a different feature. Due to the conceptual
similarity of the PAC to the notion of absolute 
residual in regression, such a plot is called a 
\textit{quasi residual plot} (QRP)
\cite{Raymaekers:Silhouettes}.

\begin{figure}[!ht]
\centering
\includegraphics[width=0.95\linewidth]
   {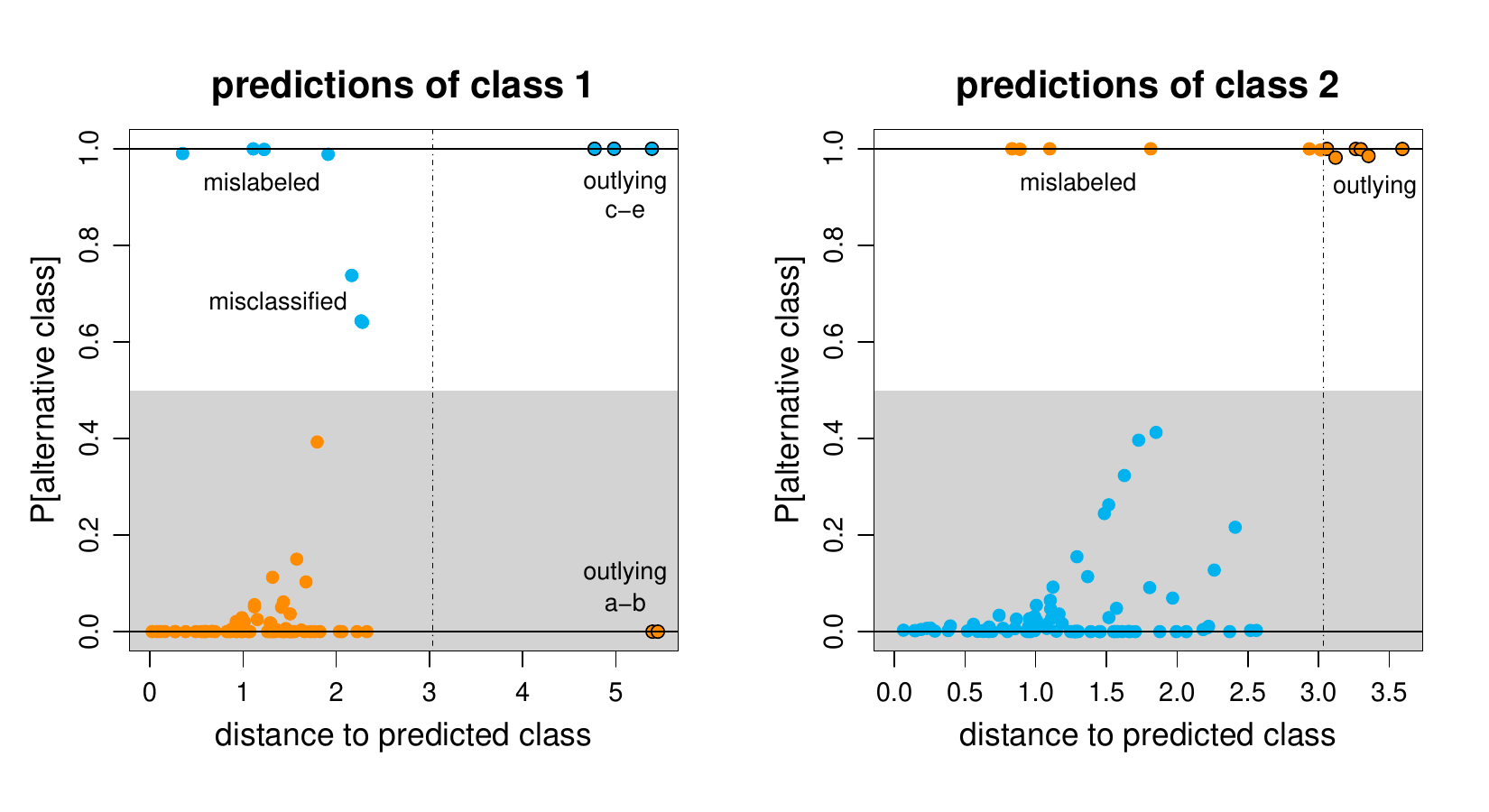} 
\vspace{-0.7cm}
\caption{Quasi residual plot with the distance to the
         predicted class on the horizontal axis.}
\label{fig:RQDA_classmap_pred}
\end{figure}

Figure~\ref{fig:RQDA_classmap_pred} shows QRPs for 
both {\it given} classes, where the feature on the 
horizontal axis is the robust distance
$\text{RD}(\bx_i,\hbmu_{\hg_i,\text{R}}, 
\hbSigma_{\hg_i,\text{R}})$ of $\bx_i$ to its 
{\it predicted} class. In these plots the colors 
of the points are those of the predicted class. 
Some points are shown with a black border. These are
the overall outliers, that is, the points whose robust
distance to all classes are above the cutoff value.

The gray zone in these plots, which is the region 
where $\PAC < 0.5$\,, contains all cases for which the 
given label matches the predicted label. 
Here both classes contain some cases above the gray
zone, so their given label does not match their 
predicted label. Most of these have a PAC close to 1, 
i.e.\ the classifier is very convinced that they should 
belong to the other class. This could be a sign of 
either mislabeling or outlyingness. Fortunately the 
horizontal dimension of the QRP provides further 
insight. The points in the top left quarter of such a
plot are assigned to a different class, and also lie 
very close to this predicted class. This is an 
indication of mislabeling. 
The points in the top right quarter of such a plot 
are cases which are assigned to a different class, 
but nevertheless have a large distance to this 
predicted class. These cases are likely to be outliers. 

The gray zone contains all the points whose given label 
agrees with the classifier. The points in the left 
part lie close to their given class, so for them 
everything seems fine. The points in the bottom right 
lie far from their class. Cases \texttt{a} and 
\texttt{b} are examples of such outliers, and in 
Figure~\ref{fig:noisyRQDA} they are indeed outliers 
of class~1, but they do not lie in the direction 
of class~2.

Instead of plotting the distance of a case to 
its predicted class on the horizontal axis as 
in Figure~\ref{fig:RQDA_classmap_pred}, one can 
also plot the distance 
$\text{RD}(\bx_i,\hbmu_{g_i,\text{R}}, 
\hbSigma_{g_i,\text{R}})$
to its {\it given} class on the horizontal axis. 
In \cite{Raymaekers:ClassMap} a related quantity
called {\it farness} is plotted, which is more 
general in the sense that it can be extended to 
several other classifiers. For discriminant
analysis, it is based on an estimate of the 
cumulative distribution function of this distance 
for random cases $\bX$ generated from class $g$. 
The farness is then defined as
\begin{equation}\label{eq:farness}
    \mbox{farness}(\bx_i) =
    P(\text{RD}(\bX,\hbmu_{g_i,\text{R}}, 
    \hbSigma_{g_i,\text{R}})\leqslant
    \text{RD}(\bx_i,\hbmu_{g_i,\text{R}}, 
    \hbSigma_{g_i,\text{R}})).
\end{equation}
The farness is a probability so it lies between 0 and 1, 
and expresses how far a case is from its given class 
as the probability of a random case of the same class 
lying closer to it. Overall outliers can now be defined 
as those cases whose farness to all classes exceeds 
0.99 (or another high probability value chosen by 
the user). 
Using the farness as the quantity on the horizontal axis 
of a quasi residual plot yields the \textit{class map} \cite{Raymaekers:ClassMap}. As before, the cases are 
colored according to their predicted label, and overall 
outliers (based on the farness) have a black border.

\begin{figure}[!ht]
\centering
\includegraphics[width=0.95\linewidth]
   {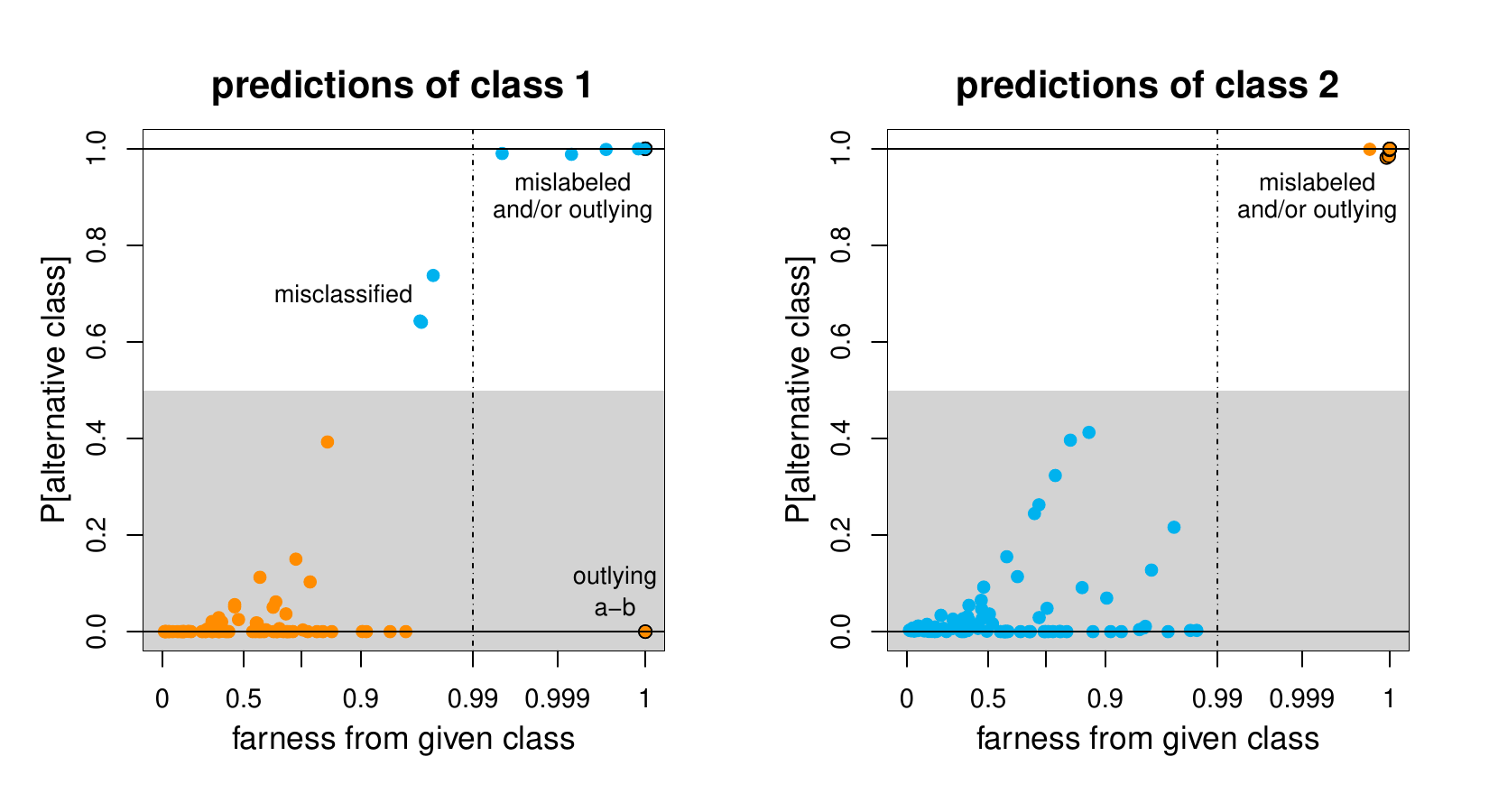} 
\vspace{-0.7cm}
\caption{Class maps of the simulated data of 
         Figure~\ref{fig:noisyRQDA}.}
\label{fig:RQDA_classmap_orig}
\end{figure}

The left panel of Figure~\ref{fig:RQDA_classmap_orig} shows 
the class map of class 1, produced by the \texttt{R}
package \texttt{classmap} \cite{classmap_R}. It plots a 
probability versus a probability, but the farness tickmarks 
on the horizontal axis are plotted at the quantiles of the 
standard normal distribution restricted to the interval 
$[0,4]$, which allows us to better distinguish between 
high farness values. The class map is similar in spirit 
to the previous QRP, but now it shows the 
distance to the {\it given} class instead of to the 
{\it predicted} class. Most of the suspicious points now 
appear in the top right quadrant of the class map. The 
true outliers have a farness close to 1, indicating that 
it is virtually impossible to generate a clean case from 
the given class that lies as far from the center as these 
points. The cases that were previously found to be 
mislabeled now \red{also} appear in the top right portion of 
the plot.

\section*{\sffamily \Large EXAMPLE}

We illustrate RQDA and the graphical displays on a 
6-dimensional real data set about floral pear buds,
with 4 classes \cite{Wouters:floral}. The goal is 
to classify the 550 cases into buds, branches, 
bud scales and support.

\begin{figure}[!ht]
\centering
\includegraphics{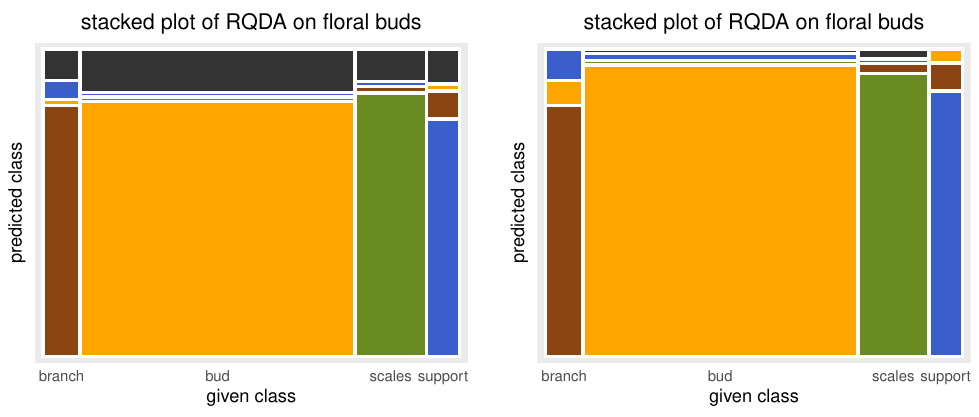}
\caption{Stacked plots of the floral buds data 
where the outlier class is defined (left) through 
robust distances and (right) through farness.}
\label{fig:floral_mosaic}
\end{figure}

Applying RQDA to this data set yields the 
stacked mosaic plot displayed in the left panel of Figure~\ref{fig:floral_mosaic}.
We see at a glance that the buds class is by far the 
largest group, with few misclassified cases. But it 
contains many cases that are classified as overall 
outliers (52 out of 363). Branch and support form 
smaller, yet partially overlapping classes.

\begin{figure}[!ht]
\centering
\includegraphics[width=0.9\linewidth]
   {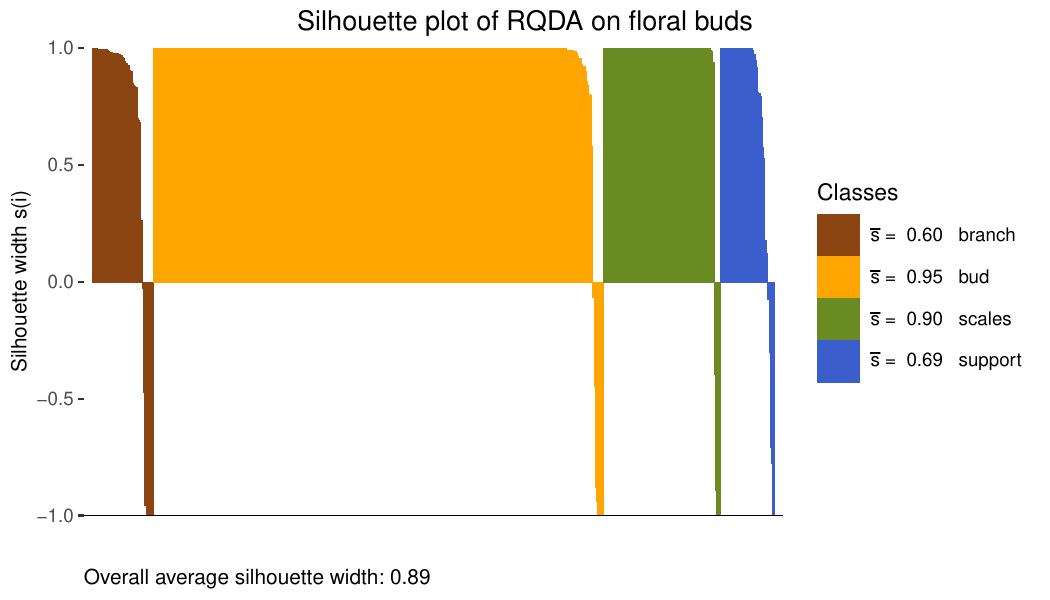} 
\vspace{-0.3cm}
\caption{Silhouette plot of the floral bud data.}
\label{fig:floral_silplot}
\end{figure}

The resulting silhouette plot is in
Figure~\ref{fig:floral_silplot}.
(Here the silhouette widths are displayed on 
the vertical axis, which can be more 
convenient on a screen in landscape mode.) 
We see that the average silhouette widths of 
classes branch and support are lower than those 
of bud and scales. The branch class has only a 
few cases that fit perfectly within their 
given group. 

\begin{figure}[!ht]
\centering
\includegraphics[width=0.95\linewidth]
   {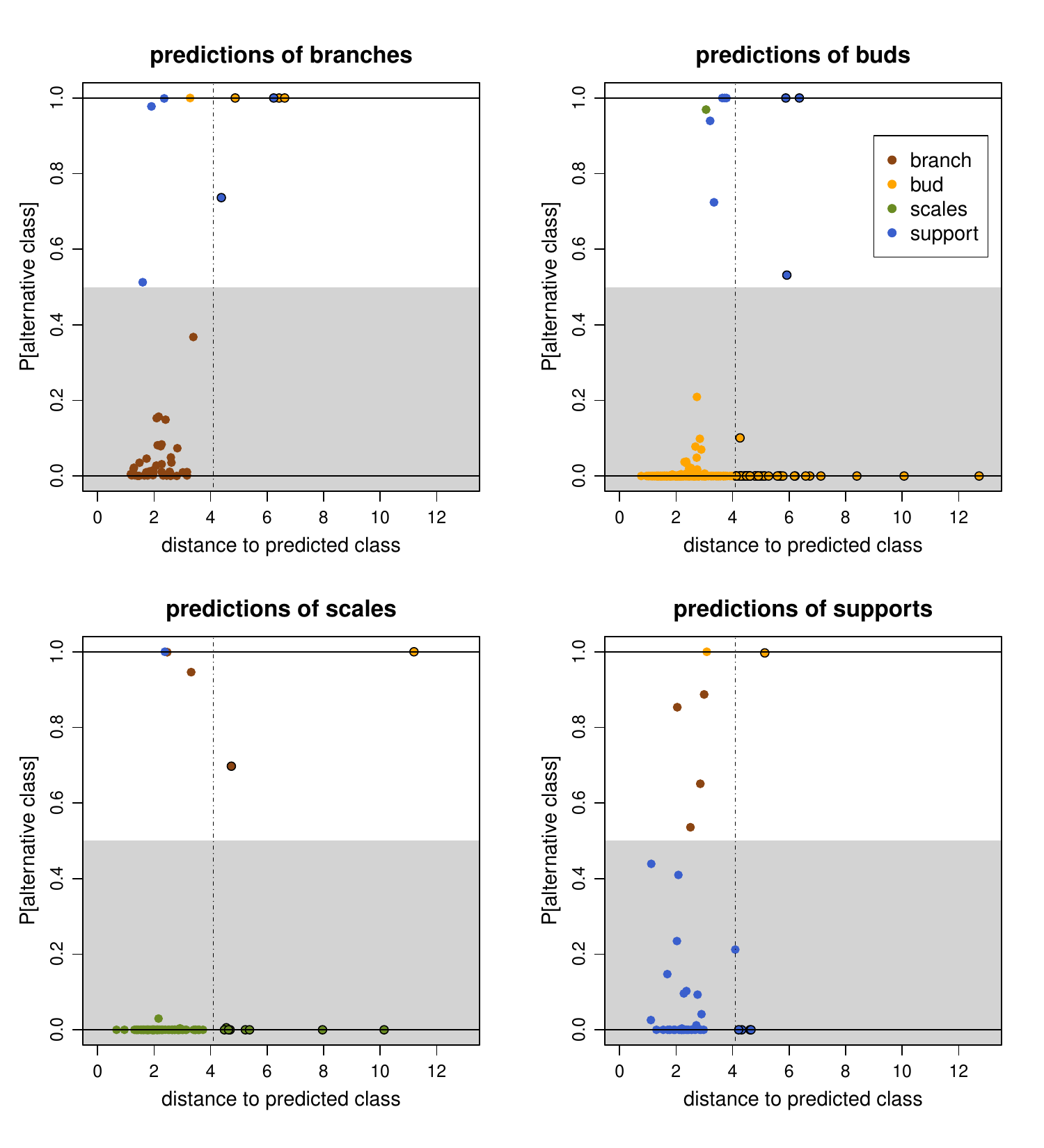} 
\vspace{-0.2cm}
\caption{Quasi residual plots of the floral bud 
data, with the distance to the predicted class 
on the horizontal axis.}
\label{fig:floral_qrp_distance}
\end{figure}

The quasi residual plots in 
Figure~\ref{fig:floral_qrp_distance} provide 
additional information. Looking at the distance to 
the predicted class on the abscissa, one notes that 
many cases with large distance have a black border, 
meaning they are distant from all groups. 
The PAC values of the support class are spread out, 
whereas most of the distant cases within buds and
scales are classified into their given group
anyway. 

\begin{figure}[!ht]
\centering
\includegraphics[width=0.98\linewidth]
   {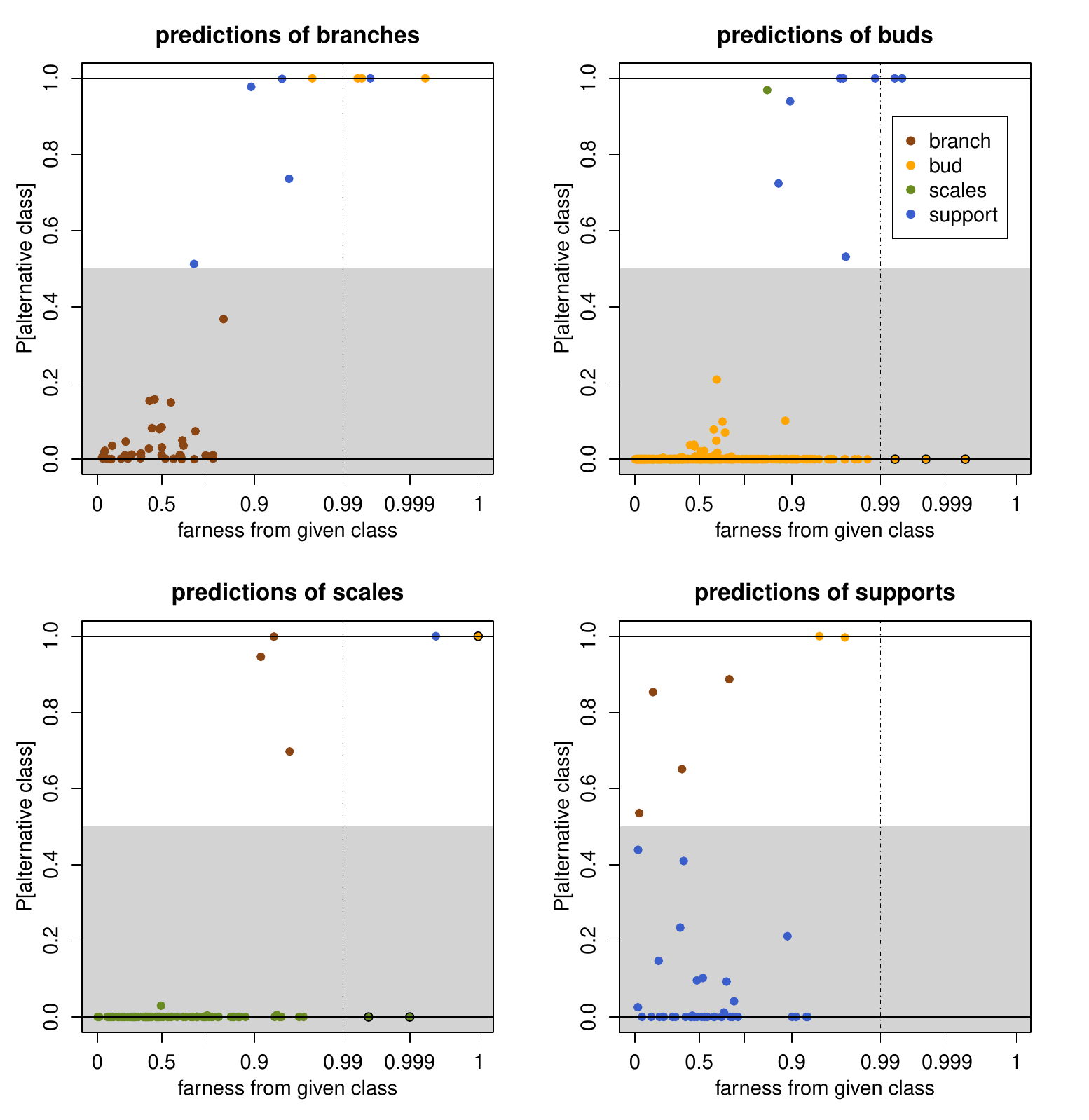} 
\vspace{-0.2cm}
\caption{Class maps of the floral bud data.}
\label{fig:floral_classmap}
\end{figure}

The original class maps are shown in
Figure~\ref{fig:floral_classmap}. Now the
horizontal axis shows the farness of each case
instead of its robust distance. In this figure
the classes bud and scale have much fewer 
outliers in the horizontal direction than in 
Figure~\ref{fig:floral_qrp_distance}.

\begin{figure}[!ht]
\centering
\includegraphics[width=0.5\linewidth]
   {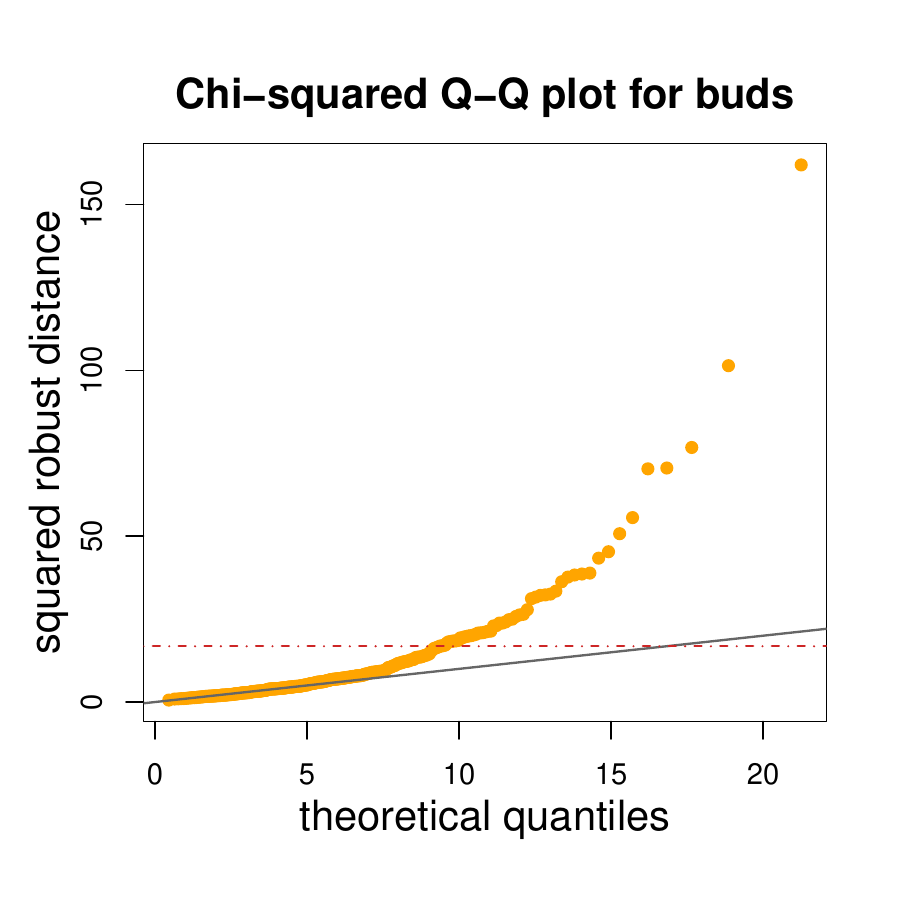} 
\vspace{-0.5cm}
\caption{Chi-squared Q-Q plot of the squared robust 
distances of the bud class. The gray line indicates
their ideal relation under normality.}
\label{fig:floral_qqplot}
\end{figure}

To understand this difference, look at the 
plot of the squared robust distances of the buds 
versus quantiles of the $\chi^2_6$ distribution 
in Figure~\ref{fig:floral_qqplot}. If the buds 
truly followed a normal distribution, the distances
would roughly follow the $\chi^2_6$ distribution, so
the points in this plot would lie close to the 
gray straight line in the plot. Instead the 
distances go up much faster, 
so their distribution has a much longer right 
tail. Therefore many squared distances exceed the 
$\chi^2_{6, 0.99}$ cutoff indicated by the red 
dashed-dotted line. On the other hand, the 0.99 
cutoff on farness is based on a robust estimate of 
the CDF of the \textit{observed} robust distances, 
and yields much fewer outliers. The stacked plot 
where the outlier class is determined by farness 
can be seen in the right panel of 
Figure~\ref{fig:floral_mosaic}, and has much smaller 
black panels. This illustrates that farness
adapts itself better to non-normally distributed 
classes.

\begin{figure}[!ht]
\centering
\includegraphics[width=0.6\linewidth]
   {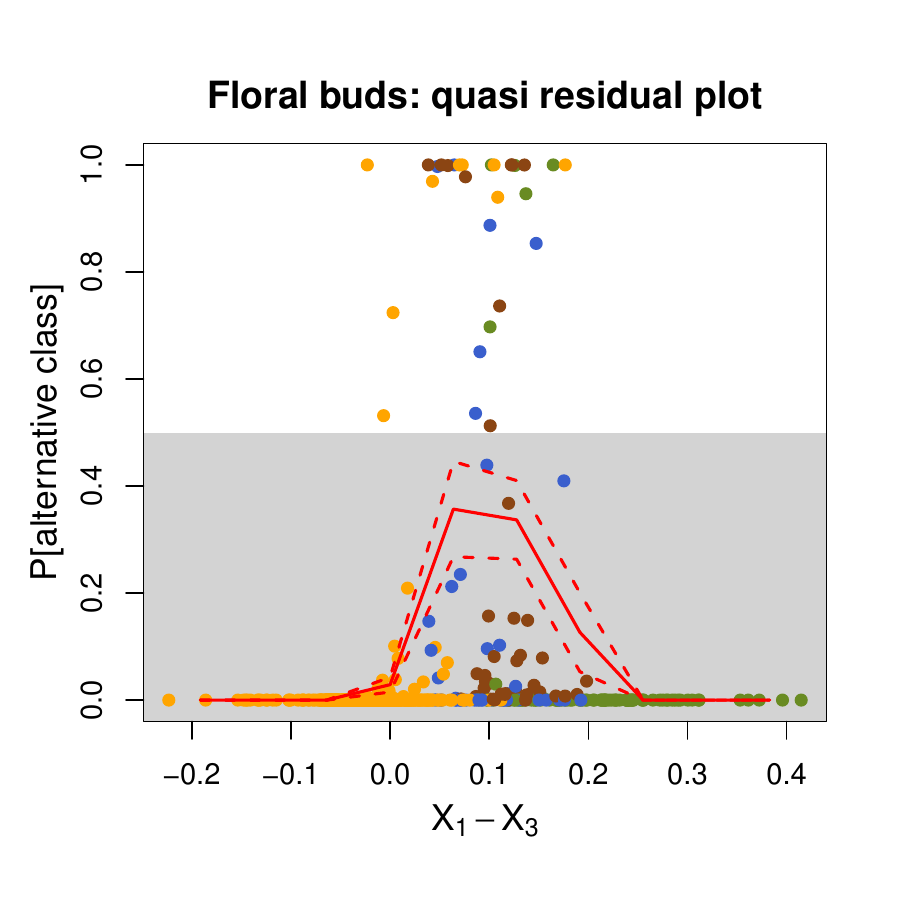} 
\vspace{-0.5cm}
\caption{Quasi residual plot of the floral bud data 
with the difference between variables 1 and 3 as 
feature on the horizontal axis. The solid red line 
represents the average PAC, and the dashed lines
show the average PAC plus or minus one standard 
error.}
\label{fig:floral_qrp}
\end{figure}

Figure~\ref{fig:floral_qrp} provides an example 
of a quasi residual plot with another feature 
on the horizontal axis, namely the difference 
between the first and third coordinate. All cases 
are shown together in a single plot, and colored 
according to their given label. The average 
PAC is the solid red curve, and the dashed curves 
correspond to the average plus or minus one 
standard error. This quasi residual plot illustrates 
that the classification works very well for small
and large values on the horizontal axis, which are 
for the most part attained by the bud and scale 
classes. The branch and support classes have 
intermediate values for this feature, and often 
obtain higher PAC values. These two classes are 
indeed not well separated.

\section*{\sffamily \Large EXTENSIONS}

So far the focus was on outliers of the casewise 
type, also called rowwise outliers because they 
correspond to a row of the $n \times p$ data 
matrix. In recent times also {\it cellwise 
outliers} are being considered. These are 
suspicious cells (entries) that can occur anywhere 
in the data matrix. \red{If $p$ is high,} even a relatively small 
proportion of outlying cells can contaminate over 
half the rows, which is a problem for existing 
rowwise robust methods. For a survey of 
the challenges posed by cellwise outliers
see \cite{Challenges}. Recently a cellwise robust
version of the MCD estimator was devised
\cite{cellMCD}. This method and other techniques
for dealing with cellwise outliers and missing
values are available in the \textsf{R} package 
\texttt{cellWise} \cite{cellWise}.

Cellwise outliers can occur in classification
data. A cellwise robust version of linear 
and quadratic discriminant analysis has been
constructed \cite{Aerts2017}. To this end 
cellwise robust estimates of the location and
covariance matrix of each class are computed,
and plugged into the assignment rule. Further 
work could deal with the situation that new 
data points to be classified may also contain
outlying cells.

In functional data analysis, a case
is not a row of a matrix but a function,  
such as a curve. 
In that setting, cellwise outliers
correspond to local deviations (like
spikes) in a curve, rather than global
outlyingness of the entire curve.
These are called {\it isolated outliers}. 
Classification methods for functional data 
with such outliers have been
constructed \cite{Hubert:MFC}.

The PAC and farness measures are not
restricted to discriminant analysis.
They have also been developed for
$k$-nearest neighbors, support vector
machines, and logistic regression
\cite{Raymaekers:ClassMap}, as well as
for CART, random forests, and neural nets
\cite{Raymaekers:Silhouettes}. For all 
these classifiers one can display silhouettes, 
quasi residual plots, and class maps, using 
the \textsf{R} package \texttt{classmap}
\cite{classmap_R}.

\section*{\sffamily \Large CONCLUSIONS}

Discriminant analysis remains one of the most 
popular methods for classification. Classical 
discriminant analysis (CDA) assumes that each 
class consists of cases that follow a 
multivariate normal distribution. It estimates 
the location and scatter matrix of each class 
by the arithmetic mean and sample covariance 
matrix. CDA therefore inherits the outlier
sensitivity of these estimators. We have 
illustrated the sensitivity of CDA to outliers 
as well as mislabeled cases through a synthetic 
data example, where its decision boundary was 
substantially distorted by the presence of a 
few outliers. This weakness makes CDA 
unreliable in many real-world settings where 
data often contains suspicious cases. 

By substituting robust estimators of location 
and scatter in the computation of discriminant 
scores, a version of DA is obtained which 
inherits the robustness properties of the 
plug-in estimator. The minimum covariance 
determinant estimator (MCD) has often been 
used for this purpose, and the robustness of 
the resulting procedure was illustrated on 
the synthetic dataset.

When a dataset contains suspicious cases, it 
may not suffice to robustly fit a classification 
rule. While this guarantees reliable estimates 
of the model parameters, it does not give us any 
insight into the nature and number of suspicious 
cases. As these cases may contain valuable 
information, they should be identified. We 
surveyed several graphical tools that can help 
with identifying outliers and mislabeled cases 
in discriminant analysis. A central role in these 
visualizations is played by the probability of 
the best alternative class (PAC). The PAC of a 
case provides a continuous measure which ranges 
from perfectly classified to misclassified. Cases 
with PAC close to zero are assigned to their 
given class with high conviction. For cases with 
PAC close to one, the prediction disagrees 
strongly with their given class. In this sense, 
the PAC provides a counterpart to the absolute 
residual in regression analysis. The PAC can 
be used to draw silhouette plots, which 
provide a quick overview of the quality of the 
fit on each class and on the entire dataset. 
We also showed several quasi residual plots,
which plotted the PAC against the distance of 
each case from its predicted class, as well as 
against its farness from its given class. 
The resulting diagnostic plots help to 
distinguish between measurement outliers and 
mislabeled cases.

\section*{\sffamily \Large CONFLICT OF INTEREST}

The authors declare that they have no conflict of interest. 


\end{document}